\documentclass[aps,prl,reprint,superscriptaddress]{revtex4-1}

\usepackage{graphicx,epsfig}
\usepackage{amsmath,amssymb,bm}
\usepackage{color}
\usepackage[usenames,dvipsnames]{xcolor}
\usepackage[colorlinks=true,linkcolor=Maroon,citecolor=OliveGreen,urlcolor=Blue,linktoc=page]{hyperref}
\usepackage{placeins}

\begin{document}


\title{
Hidden Fermi-liquid charge transport in the antiferromagnetic phase of the electron-doped cuprates
}


\author{Yangmu~Li}
\affiliation{School of Physics and Astronomy, University of Minnesota, Minneapolis, Minnesota 55455, USA}
\author{W.~Tabis}
\thanks{Current address: Laboratoire National des Champs Magn\'etiques Intenses, CNRS UPR-3228, 31400 Toulouse, France}
\affiliation{School of Physics and Astronomy, University of Minnesota, Minneapolis, Minnesota 55455, USA}
\affiliation{AGH University of Science and Technology, Faculty of Physics and Applied Computer Science, 30-059 Krakow, Poland}
\author{G.~Yu}
\affiliation{School of Physics and Astronomy, University of Minnesota, Minneapolis, Minnesota 55455, USA}
\author{N.~Bari\v{s}i\'c}\email[]{barisic@ifp.tuwien.ac.at}
\affiliation{Fakult\"at f\"ur Physik, Technische Universit\"at Wien, Wiedner Hauptstra$\ss$e 8, 1040 Wien, Austria}
\affiliation{Department of Physics, Faculty of Science, University of Zagreb, HR-10000 Zagreb, Croatia}
\author{M.~Greven}\email[]{greven@physics.umn.edu}
\affiliation{School of Physics and Astronomy, University of Minnesota, Minneapolis, Minnesota 55455, USA}

\date{\today}

\date{\today}
\begin{abstract} 
Systematic analysis of the planar resistivity, Hall effect and cotangent of the Hall
angle for the electron-doped cuprates reveals underlying Fermi-liquid behavior even deep in
the antiferromagnetic part of the phase diagram. The transport scattering rate exhibits a quadratic
temperature dependence, and is nearly independent of doping, compound and carrier type (electrons vs. holes), 
and hence universal. Our analysis moreover indicates that the material-specific
resistivity upturn at low temperatures and low doping has the same origin in both electron- and
hole-doped cuprates.
\end{abstract}



\maketitle

The cuprates feature a complex phase diagram that is asymmetric upon electron- versus hole-doping \cite{ArmitageRMP10}
and plagued by compound-specific features associated with different types of disorder and crystal structures \cite{EisakiPRB04},
often rendering it difficult to discern universal from non-universal properties. What is known for certain is that the parent compounds 
are antiferromagnetic (AF) insulators, that AF correlations are more robust against doping with electrons than with holes \cite{KeimerPRB92, MotoyamaNature07}, 
and that pseudogap (PG) phenomena, seemingly unusual charge transport behavior, and $d$-wave superconductivity appear upon doping the quintessential CuO$_2$ 
planes \cite{ArmitageRMP10}. The nature of the metallic state that emerges upon doping the insulating parent compounds has remained a central open question. 
Moreover, below a compound specific doping level, the low-temperature resistivity for both types of cuprates develops a logarithmic
upturn that appears to be related to disorder, yet whose microscopic origin has remained unknown \cite{ArmitageRMP10, RadhakrishnanPHYC90, AndoPRL95, Rullier-AlbenqueEPL08}. 
In contrast, at high dopant concentrations, the cuprates are good metals with well-defined Fermi surfaces and clear evidence for Fermi-liquid (FL) behavior
\cite{YoshidaPRB01,NakamaePRB03,HusseyNature03,VignolleNature08,NakamaePRB09,HelmPRL09, HelmPRL10}.

In a new development, the hole-doped cuprates were found to exhibit FL properties in an extended temperature range below the characteristic 
temperature $T^{**}$ ($T^{**} < T^*$; $T^*$ is the PG temperature): (i) the resistivity per CuO$_2$ sheet exhibits a universal, quadratic temperature dependence, and is inversely 
proportional to the doped carrier density $p$, $\rho \propto T^2/p$ \cite{BarisicPNAS13}; (ii) Kohler's rule for the magnetoresistvity, the characteristic of a 
conventional metal with a single relaxation rate, is obeyed, with a Fermi-liquid scattering rate, $1/\tau \propto T^2$ \cite{ChanPRL14}; (iii) the
optical scattering rate exhibits the quadratic frequency dependence and the temperature-frequency scaling expected for a Fermi liquid \cite{MirzaeiPNAS13}. 
In this part of the phase diagram, the Hall coefficient is known to be approximately independent of temperature and to take on a value that corresponds to $p$, 
$R_{\mathrm{H}} \propto 1/p$ \cite{AndoPRL04a}.

In order to explore the possible connection among the different regions of the phase diagram, an important quantity
to consider is the cotangent of the Hall angle, $\cot(\theta_{\mathrm{H}})=\rho/(HR_{\mathrm{H}})$. For simple metals, this quantity
is proportional to the transport scattering rate, $\cot(\theta_{\mathrm{H}}) \propto m^*/\tau$ ($H$ the magnetic field, and $m^*$ the effective mass).
It has long been known that $\cot(\theta_{\mathrm{H}}) \propto T^2$ in the ``strange-metal" (SM) regime ($T > T^*$) of the hole-doped cuprates
\cite{ChienPRL91}, where $\rho \propto T$ \cite{ArmitageRMP10}, which has been interpreted as the result of distinct longitudinal and transversal scattering rates \cite{AndersonPRL91} or due to an anisotropic scattering rate \cite{HusseyJPCM08,BuhmannPRB13}. Remarkably, for the model compound HgBa$_2$CuO$_{4+\delta}$ (Hg1201) it was recently found that  
$\cot(\theta_{\mathrm{H}})\propto T^2$ is independent of doping and does not exhibit a noticeable change upon crossing the characteristic temperatures $T^*$ and $T^{**}$, thus
providing a direct link between the SM and PG/FL regimes \cite{BarisicPreprint15}. Upon combining this result with those
for other hole-doped cuprates, it was furthermore shown that the transport scattering rate is
approximately compound independent, and hence that the scattering mechanism characteristic of the FL at high  
doping levels ($p \approx 0.3$) prevails even at $p=0.01$, very close to the Mott-insulating state \cite{BarisicPreprint15}.

%

\begin{figure*}[t!]
\begin{center}
\includegraphics[width=16.5cm]{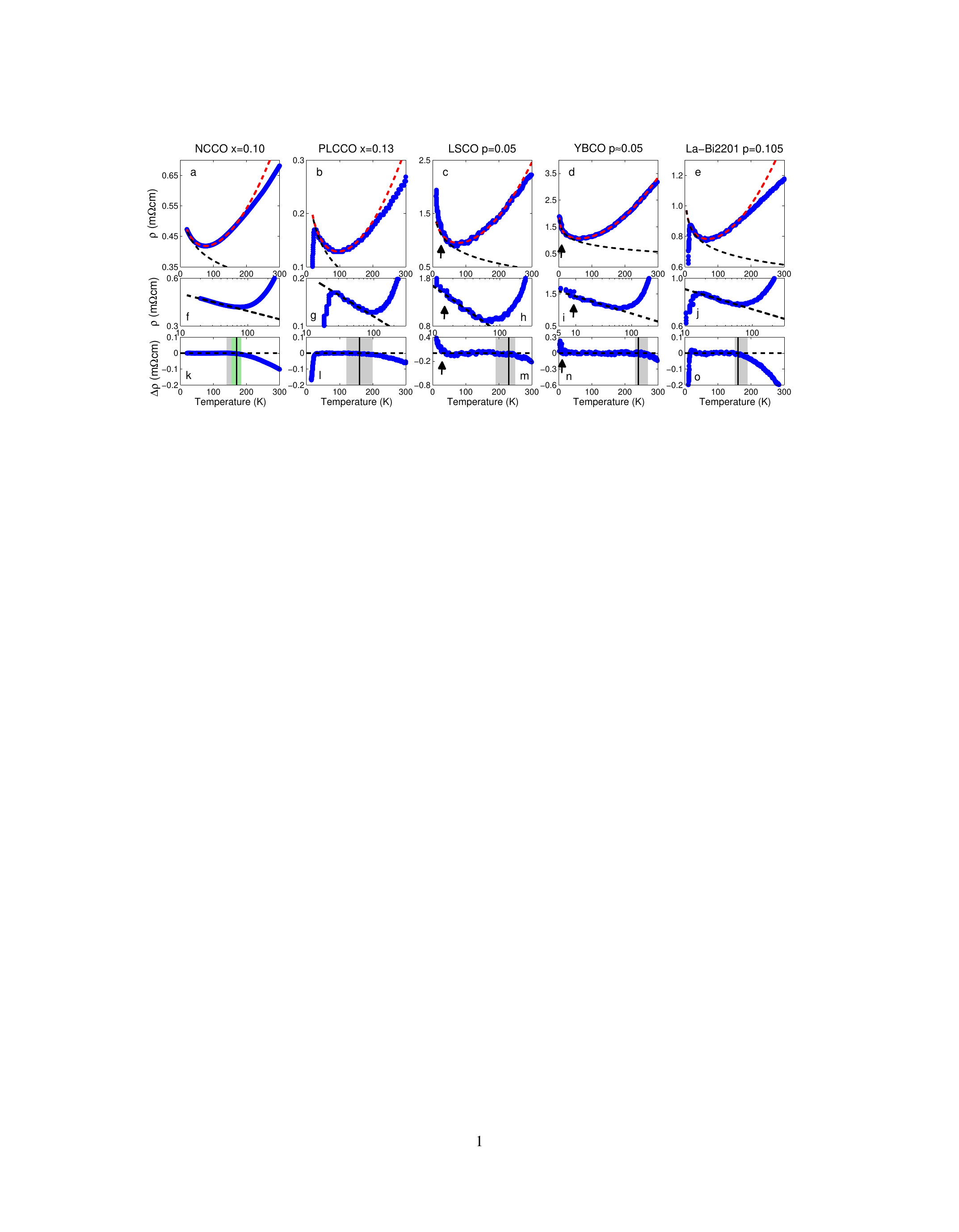}
\end{center}
\vspace{-7mm}
\caption{$ab$-plane resistivity of various cuprate materials. (a-e) Raw data (blue circles) and fit to Eq.~\ref{eqn.3} (red dashed curves). The estimated contributions $A_0-A_{\mathrm{log}}\log(T/1\mathrm{K})$ are shown as black dashed lines. (f-j) Semilog plots of the resistivity. The dashed lines indicate the logarithmic contribution. (k-o) Differences between raw data and fits. Horizontal black dashed lines indicate zero difference and are guides to the eye. Black vertical lines indicate the temperatures above which the fits deviate from the data. Grey shaded bands indicate the temperature range in which the underlying quadratic temperature dependence of the planar resistivity breaks down \cite{AndoPRL042, SupplementNCCOPRL}. The N\'{e}el temperature ($T_\mathrm{N}=165 \mathrm{K}$) of NCCO ($x=0.10$) is shown as a green shaded band \cite{MotoyamaNature07}. Arrows indicate low-temperature deviations from logarithmic behavior in lightly doped LSCO and YBCO. Except for the new NCCO data, which are consistent with prior work \cite{OnosePRB04}, data adapted from \cite{AndoPRL042, PengNatComm13, SunPRL04}.}
\vspace{-4mm}
\label{figure1}
\end{figure*}

%

Here we consider the electron-doped half of the phase diagram. Unlike for the hole-doped cuprates \cite{BarisicPreprint15}, 
we find it necessary to explicitly consider the low-T logarithmic upturn of the resistivity, $\Delta\rho(T)\propto -\log(T)$. The magnitude of this upturn is 
non-universal, can vary from sample to sample for the same compound and doping level, and is particularly large deep in the AF state of
the electron-doped compounds. This analysis reveals underlying FL behavior in the AF state. Moreover, the transport scattering rate is nearly 
the same as for the hole-doped cuprates. Furthermore, these surprising new insights allow us to extend the prior analysis of hole-doped compounds to lower temperature 
and to demonstrate that the resistivity upturn must have the same physical origin in both electron- and hole-doped cuprates.

It is instructive to recall the systematic study of initially very clean, hole-doped YBa$_{2}$Cu$_{3}$O$_{6+\delta}$ (YBCO) samples, with intrinsic resistivity $\rho_{i}(T)  \propto T^2$ in the PG regime \cite{BarisicPNAS13},
that were subsequently exposed to electron-beam irradiation \cite{Rullier-AlbenqueEPL08}. Upon increasing the radiation dose, and hence the density of point defects, the resistivity was found to be enhanced by a T-independent contribution ($\rho_0$) and a low-T upturn ($\Delta\rho(T)$). 
This suggests that the resistivity can be decomposed into three terms: 
\begin{equation} \label{eqn.1}
\rho=\rho_{0}+\Delta\rho(T)+\rho_{i}(T)\tag{1}
\end{equation}
Except at very low doping levels and temperatures, the non-universal upturn is known to exhibit a logarithmic temperature dependence \cite{Rullier-AlbenqueEPL08, AndoPRL95, OnoPRL00, DoironleyraudPRL06}. 

Starting with new data for a sample of the archetypal electron-doped cuprate Nd$_{2-x}$Ce$_x$CuO$_{4+\delta}$ (NCCO) that exhibits robust AF order ($x = 0.10$; superconductivity in NCCO appears for $x \approx 0.13$ \cite{MotoyamaNature07}), we follow the evolution of these three contributions as a function of temperature and doping for a large number of compounds: electron-doped NCCO \cite{OnosePRB04, WangSST05, WangPRB05, WoodsPRB98, BachPRB11}, La$_{2-x}$Y$_x$CuO$_{4+\delta}$ (LYCO) \cite{YuPRB07}, Pr$_{2-x}$Ce$_x$CuO$_{4+\delta}$ (PCCO) \cite{DaganPRL04, FinkelmanPRB10, GauthierPRB07, TaftiPRB14}, Pr$_{1.3-x}$La$_{0.7}$Ce$_x$CuO$_{4+\delta}$ (PLCCO) \cite{SunPRL04}, and La$_{2-x}$Ce$_x$CuO$_{4+\delta}$ (LCCO) \cite{JinPRB09}, and hole-doped La$_{2-x}$Sr$_x$CuO$_4$ (LSCO) \cite{AndoPRL95, AndoPRL042}, YBCO \cite{AndoPRL042, LeePRB05, Rullier-AlbenqueEPL08} and Bi$_2S$r$_{2-x}$La$_x$Cu$_2$O$_{8+\delta}$ (La-Bi2201) \cite{OnoPRL00, PengNatComm13}. Representative resistivity data are shown in Fig.~\ref{figure1} (for a summary of sample characteristics, see \cite{SupplementNCCOPRL}). In all cases, the logarithmic contribution is apparent. 

Equation~\ref{eqn.1} can be written in two identical forms:
\begin{equation} \label{eqn.2}
\rho=\rho_{\mathrm{res}}-A_{\mathrm{log}}\log(T/T_{\mathrm{log}})+A_2T^2\tag{2a} 
\end{equation}
\begin{equation} \label{eqn.3}
\rho=A_{0}-A_{\mathrm{log}}\log(T/1\mathrm{K})+A_2T^2\tag{2b}
\end{equation}
where $\rho_{\mathrm{res}}$ is the residual ($T=0$) resistivity and $A_{0}=\rho_{\mathrm{res}}+A_{\mathrm{log}}\log(T_{\mathrm{log}}/1\mathrm{K})$. We fit the data (see Fig. 1) to the second form, as it contains three rather than four parameters. This procedure resembles that suggested in ref. \cite{Rullier-AlbenqueEPL08}, with $A_0=\rho_0$, except that we allow all three parameters to vary, i.e., we do not use the high-T data to fix $A_2$ and $\rho_0$; the difference between these two approaches, which lead to very similar conclusions, is further analyzed in \cite{SupplementNCCOPRL}. 

\begin{figure*}[t]
\hspace*{0mm}\includegraphics[height=4.3cm,width=18.5cm]{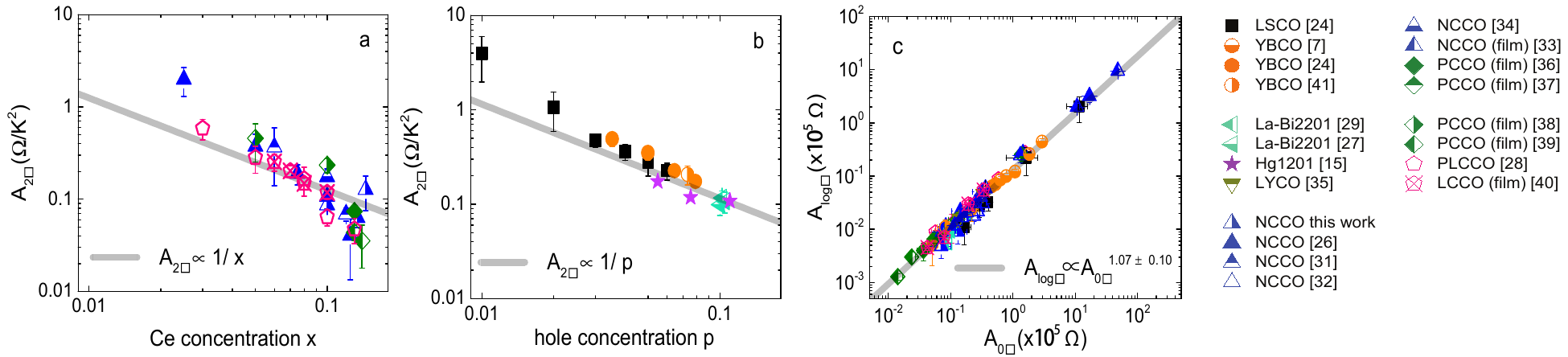}
\vspace{-6mm}
\caption{Doping dependence of $A_{2\square}$ for (a) electron- and (b) hole-doped materials, as obtained from fits to Eq.~\ref{eqn.3}. Note that in ref. \cite{BarisicPNAS13} $A_{2\square}$ was estimated for the hole-doped cuprates without considering the logarithmic contribution; see also \cite{SupplementNCCOPRL}. Grey lines are guides to the eye and indicate $1/x$ and $1/p$ dependences, respectively. 
(c) Scaling relation between $A_{\mathrm{log}\square}$ and $A_{0\square}$.
Data adapted from \cite{AndoPRL042, LeePRB05, Rullier-AlbenqueEPL08, YuPRB07, OnoPRL00, PengNatComm13, BarisicPNAS13, OnosePRB04, WangSST05, WangPRB05, BachPRB11, WoodsPRB98, DaganPRL04, FinkelmanPRB10, GauthierPRB07, TaftiPRB14, SunPRL04}.
Error bars are one standard deviation, estimated from fits with various temperature ranges. 
At low doping, the low-temperature upturn can no longer be well described by a logarithmic
contribution (see Fig. 1 and  \cite{SupplementNCCOPRL}), and the obtained coefficients thus depend on the fit temperature range and have large error bars.
}
\vspace{-4mm}
\label{figure2}
\end{figure*}

For the hole-doped cuprates, it was demonstrated that the sheet resistance coefficient $A_{2\square}$ is universal \cite{BarisicPNAS13}. Thus, to compare $A_{\mathrm{log}}$ and $A_2$ for different compounds, we convert to sheet resistance units. Figure~\ref{figure2}a shows that $A_{2\square}$ for the electron-doped materials is approximately inversely proportional to the cerium concentration, and hence to the nominal electron concentration. For comparison, Fig.~\ref{figure2}b shows $A_{2\square} \propto 1/p$ obtained from fits to Eq.~\ref{eqn.3} for the hole-doped cuprates. Remarkably, the absolute values of $A_{2\square}$ for hole- and electron-doped materials at the same nominal doping level are very similar. 

%

%

In contrast, at the same nominal doping level, $A_{\mathrm{log}\square}$ and $A_{0\square}$ are nearly an order of magnitude larger for the electron-doped compounds; similar to $A_{2\square}$, these coefficients exhibit 
power-law doping dependences \cite{SupplementNCCOPRL}: $A_{\mathrm{log}\square} \propto x^{-3.6 \pm 0.3}$ and $A_{0\square} \propto x^{-3.4 \pm 0.3}$ (electron doping) and $A_{\mathrm{log}\square} \propto p^{-3.0 \pm 0.2}$ and $A_{0\square} \propto p^{-2.7 \pm 0.3}$ (hole doping). Within error, the exponents are the same on each side of the phase diagram. 
Figure~\ref{figure2}c, which treats the doping level as an implicit parameter, highlights this point 
by revealing an approximately linear relationship 
between $A_{0\square}$ and $A_{\mathrm{log}\square}$, which holds over many orders of magnitude. 
This carries several important messages.  
First, it implies that the dominant contribution to $A_0$ is not related to residual impurity scattering (recall that $A_{0}=\rho_{\mathrm{res}}+A_{\mathrm{log}}\log(T_{\mathrm{log}}/1\mathrm{K})$). 
Second, $T_{\mathrm{log}}$ should not vary considerably.
Indeed, $T_{\mathrm{log}}$ obtained from fits to Eq.~\ref{eqn.2} is on the order of 50-150 K and exhibits a very similar monotonic doping dependence for all materials \cite{SupplementNCCOPRL}. 
Finally, these observations point to a single mechanism
that universally governs the appearance of the resistivity upturns in both electron- and hole-doped cuprates, which seems to be related not directly to a reconstruction of the Fermi surface \cite{LalibertearXiv16} but rather to the underlying disorder \cite{Rullier-AlbenqueEPL08}.

\begin{figure*}[t!]
\begin{center}
\includegraphics[width=14.8cm]{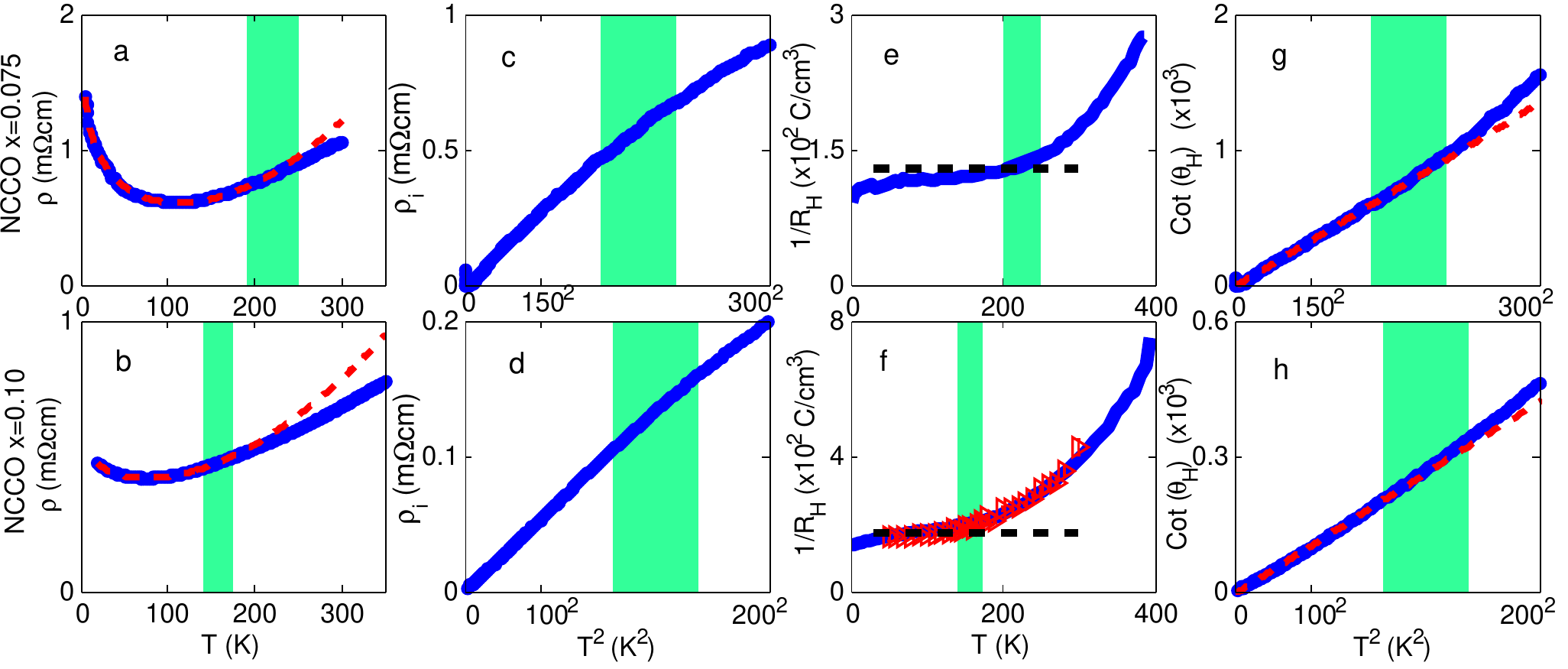}
\end{center}
\vspace{-7mm}
\caption{
Resistivity data and fit for NCCO with (a) $x=0.075$ \cite{OnosePRB04} and (b) $x=0.10$ (present work; same data as in Fig.~\ref{figure1}a). (c, d) Resistivity after subtraction of fitted logarithmic contribution. (e, f) $1/R_{\mathrm{H}}$ adapted from \cite{OnosePRL01}; $1/R_{\mathrm{H}}$ data obtained for our NCCO ($x=0.10$) sample (red triangles) agree with the prior work \cite{OnosePRL01}. Black horizontal dashed lines indicate the magnitude of $1/R_{\mathrm{H}}$ estimated from the chemical dopant concentrations $x=0.075$ and 0.10. (d, g) Cotangent of the Hall angle. Red dashed lines indicate $\cot(\theta_{\mathrm{H}})\propto C_{2}T^2$. As in Fig. 1k, green bands indicated the estimated N\'eel temperatures \cite{MotoyamaNature07}.}
\vspace{-4mm}
\label{figure3}
\end{figure*}

Motivated by these insights and by the recent finding of a universal scattering rate throughout the phase diagram of the hole-doped cuprates \cite{BarisicPreprint15} 
($\cot(\theta_{\mathrm{H}})=C_0 + C_2T^2$ holds with universal value of $C_2=0.0175(20)$ and with a compound, doping, disorder dependent $C_0$), 
we take a closer look at prior comprehensive data for the AF phase of NCCO \cite{OnosePRL01}. Figure~\ref{figure3} shows the procedure to disentangle resistivity contributions and to obtain $\cot(\theta_{\mathrm{H}})$ for
NCCO ($x=0.075$ and 0.10). The underlying Fermi-liquid scattering rate is only revealed upon considering the logarithmic upturn. (The same procedure was applied for $x=0.05$ and 0.125 \cite{OnosePRL01,OnosePRB04}; see \cite{SupplementNCCOPRL}). Once we subtract the non-universal contribution, we find
\begin{equation} \label{eqn.4}
\cot(\theta_{\mathrm{H}})=\rho_{i}/(HR_{\mathrm{H}})=C_{2}T^2\tag{3},
\end{equation}
where $\rho_{i}=\rho-(A_{0}-A_{\mathrm{log}}\log(T/1\mathrm{K}))$.

Above the N\'{e}el temperature, $\cot(\theta_{\mathrm{H}})$ deviates from this simple quadratic behavior. This is consistent with Fig.~1k and appears to be the result of Fermi-surface reconstruction \cite{OnosePRB04}: upon increasing the temperature above $T_\mathrm{N}$, the Fermi surface evolves from simple electron pockets to a more complex shape, and the Hall coefficient ceases to be a good measure of the carrier density \cite{BarisicPreprint15, OnosePRL01}. In order to address the properties of the high-temperature regime, a more elaborate analysis is required, which is beyond the scope of the present paper.

\begin{figure}[t]
\begin{center}
\vspace{0mm}\includegraphics[width=6.5cm]{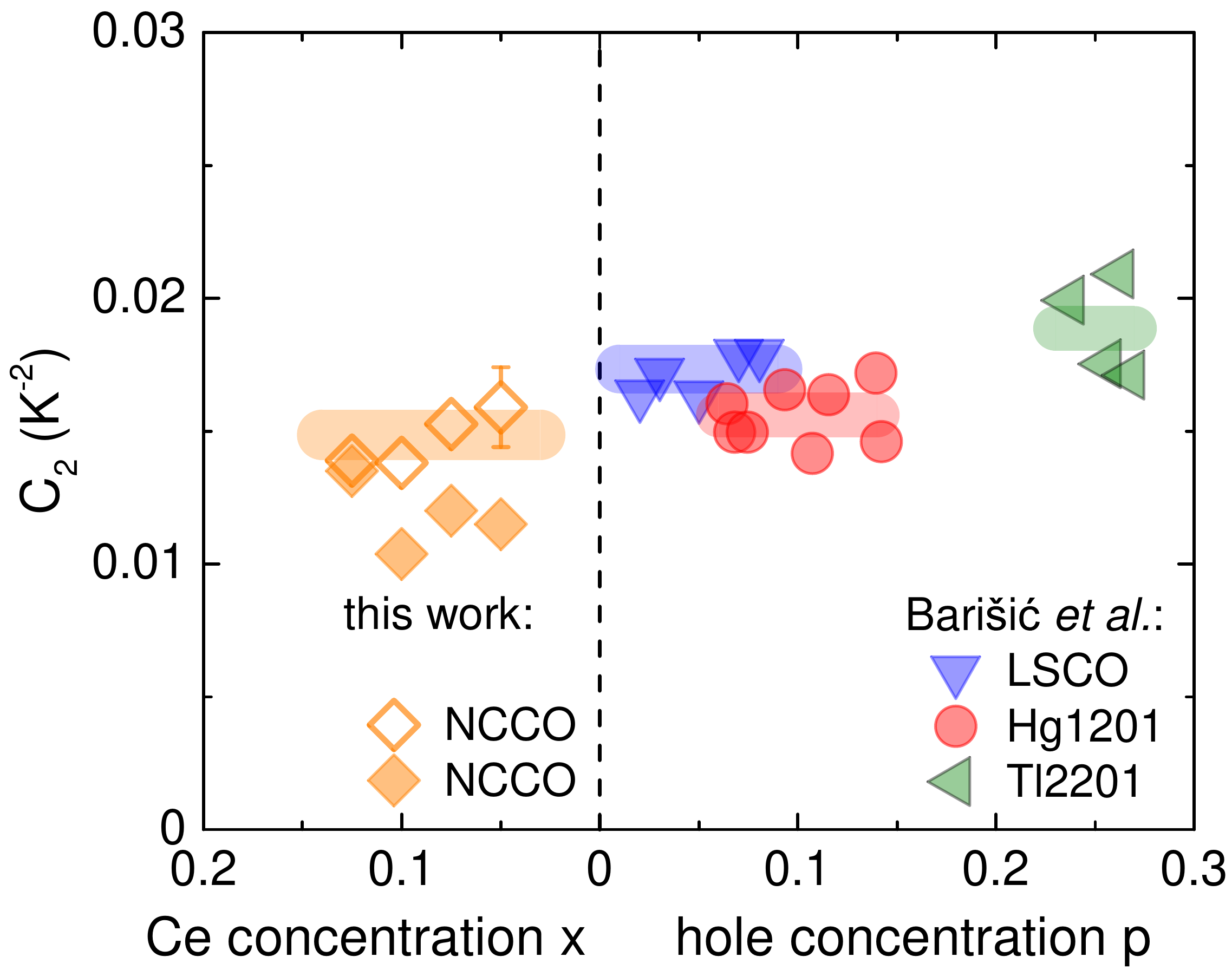}
\end{center}
\vspace{-7mm}
\caption{$C_2$ for various single-layer cuprates. LSCO, Hg1201 and Tl2201 results adapted from \cite{BarisicPreprint15}. $C_2$ for NCCO is determined both by ``method 1" (open symbols) and ``method 2" (solid symbols).
See also \cite{SupplementNCCOPRL}.}
\vspace{-6mm}
\label{figure4}
\end{figure}

As seen from Fig.~\ref{figure4}, our analysis (``method 1") of the $dc$ resistivity and Hall coefficient for NCCO yields values of $C_2$ that are nearly identical to those found previously for the hole-doped cuprates. We test the robustness of this result with regard to the fit procedure by determining the slope of the quadratic term ($A_2$) simply from the high-T part of the resistivity, as suggested in refs. \cite{Rullier-AlbenqueEPL08, BarisicPreprint15}, neglecting the low-T upturns (``method 2"). In essence, method 2 yields a lower bound for $A_2$, since the logarithmic contribution is neglected (see \cite{SupplementNCCOPRL}). As summarized in Fig.~\ref{figure4}, we find that $C_2$ changes no more than 30\% for NCCO in the studied doping range, confirming the robustness of our analysis.  

Hole-doped LSCO exhibits a considerable resistivity upturn at moderate and low doping. We have analyzed LSCO data ($p=0.05$, 0.07, 0.08) \cite{AndoPRL042} with method 1 and find hardly any change in $C_2$ compared to the prior result based on method 2 \cite{BarisicPreprint15}. However, as seen from Fig. 1c, for $p=0.05$ at the lowest temperatures, the resistivity upturn is stronger than logarithmic. Method 1 no longer gives an accurate description for $p\leq0.03$ \cite{SupplementNCCOPRL}.

The relatively small difference (about a factor of two) in the value of $C_2$ between electron- and hole-doped cuprates can be attributed to a difference in the effective mass. We are not aware of reliable measurements of $m^*$ for the electron-doped cuprates in the relevant doping range, but band-structure calculations indicate a smaller value than for the hole-doped compounds \cite{WeberNatphys10}. 

%

%
%

%

In principle, there are two distinct ways to understand the simultaneous Fermi-liquid and logarithmic transport behaviors captured by Eq.~2. The first is via Matthiessen's rule, which assumes that scattering rates for different scattering processes simply add up ($1/\tau_{total} = 1/\tau_1 +1/\tau_2 + .. $). The second possibility is that Eq.~2 describes a serial-resistor network, which would imply the existence of distinct metallic and non-metallic patches \cite{PhillipsRPP03}. We can distinguish between these two possibilities by considering a recent result obtained for the hole-doped cuprates \cite{ChanPRL14}, namely that Kohler-scaling for the orbital magnetoresistance holds for compounds/samples that exhibit negligible residual resistivity ($\rho_{\mathrm{res}} \approx 0$) and Fermi-liquid behavior ($(\rho-\rho_{H=0})/\rho_{H=0} \propto H^2/\rho_{H=0}^2 $, where $\rho_{H=0}$ is the zero field resistivity) below $T^{**}$. This scaling follows directly from the Boltzmann transport equation and unmistakably demonstrates the Fermi-liquid character of the pseudogap phase. For LSCO, which exhibits large resistivity values, 
it was found that Kohler's rule is obeyed only if $\rho_{H=0}$ is replaced by $\rho_{H=0}-A_0$. 
This surprising result is incompatible with Matthiessen's rule for a homogeneous system. However, it is compatible with a serial-resistor network \cite{HePRL12} in which only metallic patches contribute to the magnetoresistivity, whereas non-metallic regions characterized by logarithmic behavior have negligible influence. This conclusion also provides a simple explanation for the heuristic logarithmic term and for our finding (from the approximately linear scaling between $A_0$ and $A_{\mathrm{log}}$) that $\rho_{\mathrm{res}}$ is negligible. Such upturns naturally appear at sufficiently large temperatures in systems with strongly-coupled metallic grains separated by an insulating matrix \cite{BeloborodovRMP07}.

We have established that the planar charge transport of the electron-doped cuprates exhibits hidden FL behavior in the AF phase (see early discussion on FL behavior outside of the AF phase \cite{TsueiPHYC89}). This fact had previously gone unnoticed because the FL transport is masked by a particularly large non-metallic contribution to the resistivity. The FL behavior  is remarkably robust and universal. The sheet resistance coefficients $A_{2\square}$ are very similar for electron and hole-doped compounds at the same nominal doping level. Moreover, by considering the Hall effect, we demonstrate that the scattering rate is nearly the same on both sides of the phase diagram. We also find that the non-universal additive logarithmic contribution is characterized by a temperature scale that has the same magnitude for electron- and hole-doped compounds and exhibits a weak doping dependence. This fact is exemplified by the approximate (linear) scaling between $A_{\mathrm{log}\square}$ and $A_{0\square}$ that
is found to hold over four orders of magnitude. We therefore conclude that the non-metallic contribution must have the the same physical origin in all compounds. 
Most likely, it is associated with charge transport involving metallic and non-metallic regions of the material. The metallic regions continue to follow simple FL behavior 
to the lowest temperatures and doping levels studied here. Overall these insights shed new light on the physics of the quintessential copper-oxygen planes and they demonstrate that important aspects of the charge transport of electron- and hole-doped cuprates are universal.

\indent We thank E. M. Motoyama for the growth of the NCCO crystals measured in this work. We acknowledge valuable discussions with A. V. Chubukov, R. Fernandes, P. Pop\v{c}evi\'{c} and L. Forr\'o. This work was supported partially by the NSF through the University of Minnesota MRSEC under Award Number DMR-1420013, and by NSF Award 1006617. The work at the TU Wien was supported by FWF project P27980-N36.

\bibliography{bibYM2015}

\end{document}